\documentclass[twocolumn]{aastex631} 

\makeatletter

\newcommand{\Rmnum}[1]{\expandafter\@slowromancap\romannumeral #1@}
\makeatother

\newcommand{\sis}{{\ion{Si}{10}/\ion{S}{10}}}
\newcommand{\caar}{{\ion{Ca}{14}/\ion{Ar}{14}}}

\newcommand{\corr}[1]{\textcolor{black}{#1}} 

\graphicspath{{./}{figures/}}

\received{\today}
\revised{\today}
\accepted{\today}
\submitjournal{ApJ}

\shorttitle{Identifying plasma fractionation processes in the chromosphere using IRIS}
\shortauthors{Long et al.}

\begin{document}

\title{Identifying plasma fractionation processes in the chromosphere using IRIS}

\correspondingauthor{David Long}
\email{david.long@dcu.ie}

\author[0000-0003-3137-0277]{David M.~Long}
\affiliation{School of Physical Sciences, Dublin City University, Glasnevin Campus, Dublin, D09 V209, Ireland}
\affiliation{Astrophysics Research Centre, School of Mathematics and Physics, Queen’s University Belfast, University Road, Belfast, BT7 1NN, Northern Ireland, UK}

\author[0000-0002-0665-2355]{Deborah Baker}
\affiliation{Mullard Space Science Laboratory, University College London, Holmbury St Mary, Dorking, Surrey, RH5 6NT, UK}

\author[0000-0003-0774-9084]{Andy S.~H.~To}
\affiliation{ESTEC, European Space Agency, Keplerlaan 1, PO Box 299, NL-2200 AG Noordwijk, The Netherlands}

\author[0000-0002-2943-5978]{Lidia van Driel-Gesztelyi}
\affiliation{Mullard Space Science Laboratory, University College London, Holmbury St Mary, Dorking, Surrey, RH5 6NT, UK}
\affiliation{LESIA, Observatoire de Paris, Universit\'{e} PSL, CNRS, Sorbonne Universit\'{e}, Univ. Paris Diderot, Sorbonne Paris Cit\'{e}, 5 place Jules Janssen, 92195 Meudon, France}
\affiliation{Konkoly Observatory, Research Centre for Astronomy and Earth Sciences, Hungarian Academy of Sciences, Konkoly Thege \'{u}t 15-17., H-1121, Budapest,
Hungary}

\author[0000-0002-2189-9313]{David H. Brooks}
\affiliation{Department of Physics \& Astronomy, George Mason University, 4400 University Drive, Fairfax, VA 22030, USA}

\author[0000-0002-5365-7546]{Marco Stangalini}
\affiliation{ASI Italian Space Agency, Via del Politecnico, s.n.c I-00133—Roma, Italy}

\author[0000-0002-0144-2252]{Mariarita Murabito}
\affiliation{INAF - Osservatorio Astronomico di Roma, Via Frascati 33 Monteporzio Catone, Italy}
\affiliation{Space Science Data Center (SSDC) - Agenzia Spaziale Italiana, Via del Politecnico, s.n.c., I-00133, Roma, Italy}

\author[0000-0001-7927-9291]{Alexander W. James}
\affiliation{Mullard Space Science Laboratory, University College London, Holmbury St Mary, Dorking, Surrey, RH5 6NT, UK}

\author[0000-0002-7725-6296]{Mihalis Mathioudakis}
\affiliation{Astrophysics Research Centre, School of Mathematics and Physics, Queen’s University Belfast, University Road, Belfast, BT7 1NN, Northern Ireland, UK}

\author[0000-0002-0405-0668]{Paola Testa}
\affiliation{Harvard-Smithsonian Center for Astrophysics, 60 Garden Street, Cambridge, MA 02193, USA}

\begin{abstract}
The composition of the solar corona differs from that of the photosphere, with the plasma thought to fractionate in the solar chromosphere according to the First Ionisation Potential (FIP) of the different elements. This produces a FIP bias, wherein elements with a low FIP are preferentially enhanced in the corona compared to their photospheric abundance\corr{, but direct observations of this process} remain elusive. Here we use a series of spectroscopic observations of Active Region AR~12759 as it transited the solar disc over a period of 6 days from 2-7~April~2020 taken using the \emph{Hinode} Extreme ultraviolet Imaging Spectrometer (EIS) and Interface Region Imaging Spectrograph (IRIS) instruments to look for signatures of plasma fractionation in the solar chromosphere. Using the \sis\ and \caar\ diagnostics, we find distinct differences between the FIP bias of the leading and following polarities of the active region. \corr{The widths of the IRIS \ion{Si}{4} lines exhibited clear differences between the leading and following polarity regions, indicating increased unresolved wave activity in the following polarity region compared to the leading polarity region, with the chromospheric velocities derived using the \ion{Mg}{2} lines exhibiting comparable, albeit much weaker, behaviour. These results are consistent with plasma fractionation via resonant/non-resonant waves at different locations in the solar chromosphere following the ponderomotive force model, and indicate that IRIS could be used to further study this fundamental physical process.}


\end{abstract}


\section{Introduction} \label{s:intro}

Observations of elemental composition in the solar atmosphere have revealed distinct differences between the solar photosphere and corona \citep[e.g.,][]{Asplund:2009,Meyer:1985}. Elements with a low first ionisation potential (FIP) value (FIP$<10$~eV, e.g., Fe, Mg, Si) have been shown to be overabundant in the solar corona by a factor of 2--4 when compared with elements with a high FIP value (FIP $>10$~eV, e.g., C, N, O). This ratio between the composition of an element in the corona vs.\ the photosphere is known as the ``FIP bias", and provides a commonly used diagnostic of variations in abundance throughout different regions of the solar atmosphere \citep[see, e.g.,][]{Brooks:2015,Baker:2021,Laming:2015}. This is an important and fundamental property of solar plasma. Unlike other plasma properties such as temperature, density, and emission measure, which exhibit drastic changes in the corona, the FIP bias of a plasma is not affected by its surroundings. Instead, it is set low down in the solar atmosphere and does not change as the plasma evolves into the heliosphere, providing a key tool to relate remote-sensing and in-situ measurements of the solar corona and solar wind. As a result, it is a key plasma parameter observed by the \emph{Solar Orbiter} mission \citep[see, e.g.,][]{Muller:2020,Zouganelis:2020}.

However, the physical processes that drive fractionation of the plasma and produce the observed FIP bias measurements remain subject to investigation. Initial theories suggested that the observed FIP effect was due to thermal or ambipolar diffusion across magnetic field lines in the solar atmosphere \citep{vonSteiger:1989}, thermoelectric driving \citep{Antiochos:1994}, chromospheric reconnection \citep{Arge:1998}, or ion cyclotron wave heating \citep{Schwadron:1999}. While these theories successfully explain different aspects of the FIP effect, none of them can explain the observed Inverse FIP (IFIP) effect \citep{Doschek:2016}, wherein high- rather than low-FIP elements are enhanced in the corona, or low-FIP elements are depleted. More recently, a model to explain elemental fractionation and the FIP effect using the ponderomotive force was proposed by \citet{Laming:2004,Laming:2009,Laming:2015}. In this model, standing Alfv\'{e}n waves within coronal loops produce an upward directed ponderomotive force at the base of the loops (i.e., in the chromosphere), acting on ions and pulling them up into the corona. With low-FIP elements easier to ionise, this produces an increased FIP bias in closed magnetic field regions. 

The launch of the Extreme ultraviolet Imaging Spectrometer \citep[EIS;][]{Culhane:2007} onboard the \emph{Hinode} \citep{Kosugi:2007} spacecraft has enabled an unprecedented opportunity to investigate and quantify FIP bias evolution in the solar atmosphere by providing spatially resolved observations \citep[e.g.,][]{Brooks:2015,Warren:2016,Doschek:2018}. A myriad of recent \emph{Hinode}/EIS observations have also produced results consistent with the ponderomotive force model. \citet{Baker:2013,Baker:2015} used the ponderomotive force model to explain long term evolution of FIP bias in an emerging flux region within a coronal hole \citep{Baker:2013} and an active region \citep{Baker:2015}. This work was followed by \citet{Mihailescu:2022}, who found a weak dependence of FIP bias on the evolutionary stage of an active region. Despite being a static model, the ponderomotive force model has also been used by \citet{To:2021} to explain differences in the FIP bias measured using two different composition diagnostics in a small solar flare. The waves which can induce the ponderomotive force have also been observed in the chromosphere using spectropolarimetric data from the Interferometric BIdimensional Spectrometer (IBIS) instrument and related to coronal FIP bias measurements \citep{Stangalini:2021,Baker:2021}, while both \citet{Mihailescu:2023} and \citet{Murabito:2023} have noted a relationship between resonant waves and increased FIP bias via the ponderomotive force. Detections of the inverse FIP effect in the solar wind also seem to be consistent with the ponderomotive force model of abundance variations due to chromospheric fast-mode waves \citep{Brooks:2022}.

However, the fractionation process in the upper chromosphere and transition region has to date been underinvestigated. \citet{Dahlburg:2016} and \citet{Martinez-Sykora:2023} have begun the process of extending the ponderomotive force model initially proposed by \citet{Laming:2004,Laming:2009,Laming:2015} to include a multi-fluid analysis and non-equilibrium ionisation effects, both of which are important in the solar chromosphere where this process should be occurring. Observations of this region provided by the Interface Region Imaging Spectrometer \citep[IRIS;][]{depontieu:2014} spacecraft are also being used to investigate this process, despite the lack of suitable emission lines for estimating FIP bias within the wavelength range probed by IRIS. \citet{Testa:2023} tracked the evolution of an active region across a period of 10 days using observations from \emph{Hinode}/EIS, estimating the FIP bias using a novel technique employing a spectral inversion method. The derived FIP bias maps were then compared with IRIS observations processed using IRIS$^2$ \citep{Sainzdalda:2022} inversions to derive the chromospheric microturbulence. This approach suggested an enhancement of microturbulence in outflow regions exhibiting enhanced FIP bias, although no apparent relationship could be identified between microturbulence and enhanced FIP bias in an observed sunspot.

In this paper, we undertake a systematic analysis of an active region observed repeatedly over the course of 6 days by both the IRIS and \emph{Hinode} spacecraft to try and identify signatures of the plasma fractionation process in the solar chromosphere. The target active region and the different datasets used to study it are described in Section~\ref{s:obs}, with the different analysis techniques outlined in Section~\ref{s:analysis}. Section~\ref{s:events} describes the results of this analysis, with these results and their implications discussed in Section~\ref{s:disc}. Finally\corr{,} we draw conclusions and suggest potential avenues for further investigation in Section~\ref{s:conc}.

\section{Observations} \label{s:obs}

\begin{figure*}[!t]
    \centering
    \includegraphics[width=0.99\textwidth]{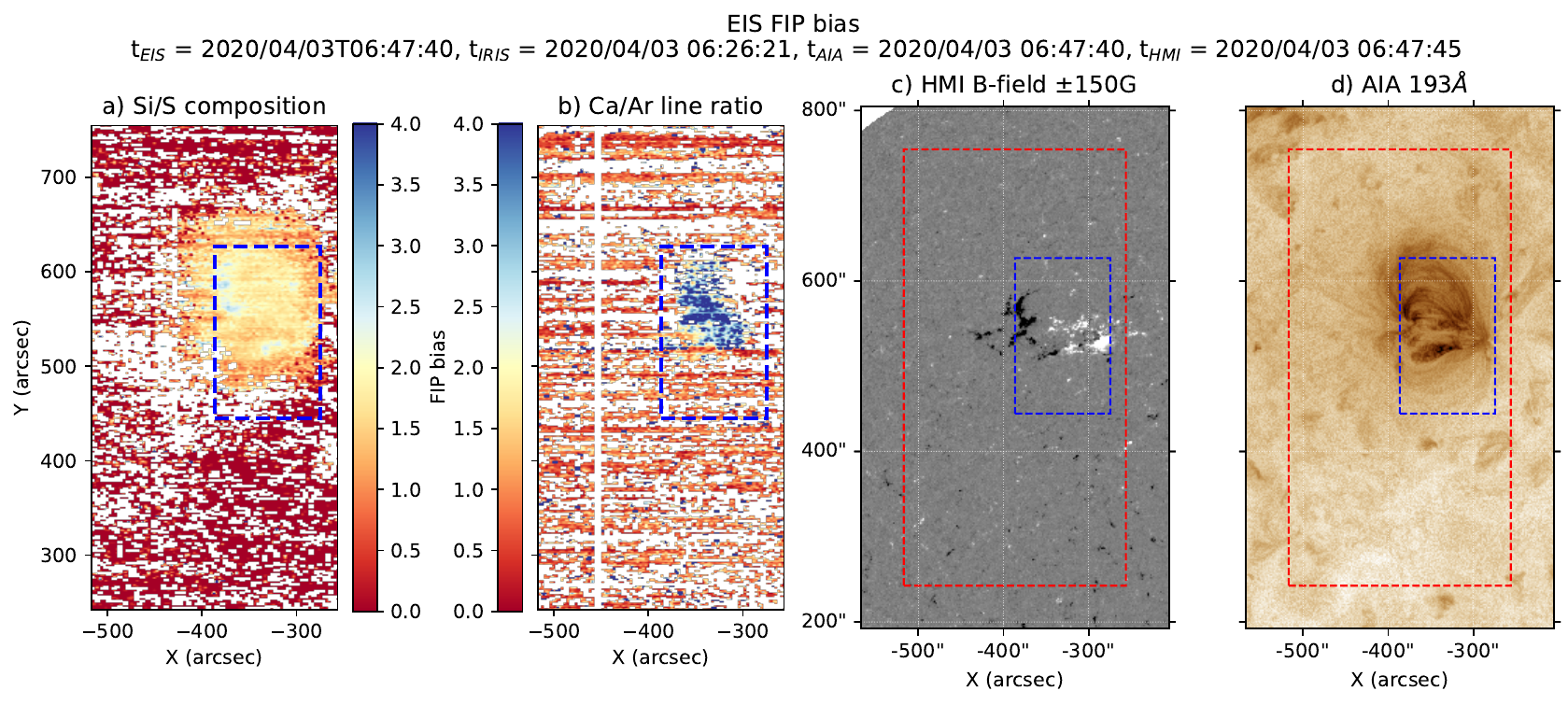}
    \caption{Active Region 12759 on 3-April-2020 observed using EIS \ion{Si}{10}/\ion{S}{10} FIP bias map (panel~a), \ion{Ca}{14}/\ion{Ar}{14} line ratio (panel~b), SDO/HMI line-of-sight magnetic field (panel~c), and SDO/AIA 171~\AA\ passband (panel~d). Magnetic field has been saturated at $\pm$150~G for clarity. White pixels in panels~a \& b denote non-numeric values. Red dashed box shows the full \emph{Hinode}/EIS field of view, and the blue dashed box shows the full IRIS field of view.}
    \label{fig:context}
\end{figure*}

The active region (AR) studied here rotated onto the solar disc as seen from Earth on 2020-March-30, and was labelled AR~12759 on 2020-April-1. At the time it was the only active region on the solar disc and appeared as a simple bipolar decaying active region (see Figure~\ref{fig:context}). As the only observable active region, it was chosen as the target for \emph{Hinode} Observing Plan (HOP) 390, with observations from both \emph{Hinode}/EIS and IRIS supporting observations made by the Karl G.~Jansky Very Large Array \citep[JVLA;][]{Perley:2011}. A more detailed discussion of the observing campaign and the relationship between the observations made using \emph{Hinode}/EIS and the JVLA can be found in \citet{To:2023}. As noted by \citet{To:2023}, the JVLA took observations of AR~12759 on 2020-April-3 and 2020-April-7, which were then used to examine the relationship between elemental abundance and F10.7 radio emission. However, as the sole active region on the disc, AR~12759 was also the focus of a series of IRIS and EIS rasters during the time period between these two JVLA observations, providing a unique insight into its long-term evolution in both the corona and the chromosphere/transition region.

The target active region was identified using observations from the \emph{Solar Dynamics Observatory} \citep[SDO;][]{Pesnell:2012}. Images from the Atmospheric Imaging Assembly \citep[AIA;][]{Lemen:2012} and both line-of-sight magnetograms and continuum images from the Helioseismic and Magnetic Imager \citep[HMI;][]{Schou:2012} were used in this analysis. These data were downloaded from the Joint Science Operations Centre (JSOC) and processed using the Python aiapy package \citep{Barnes:2020} to update the pointing, co-register the images, correct for degradation, and normalise the exposure time.

\begin{figure*}[!t]
    \includegraphics[width=\textwidth]{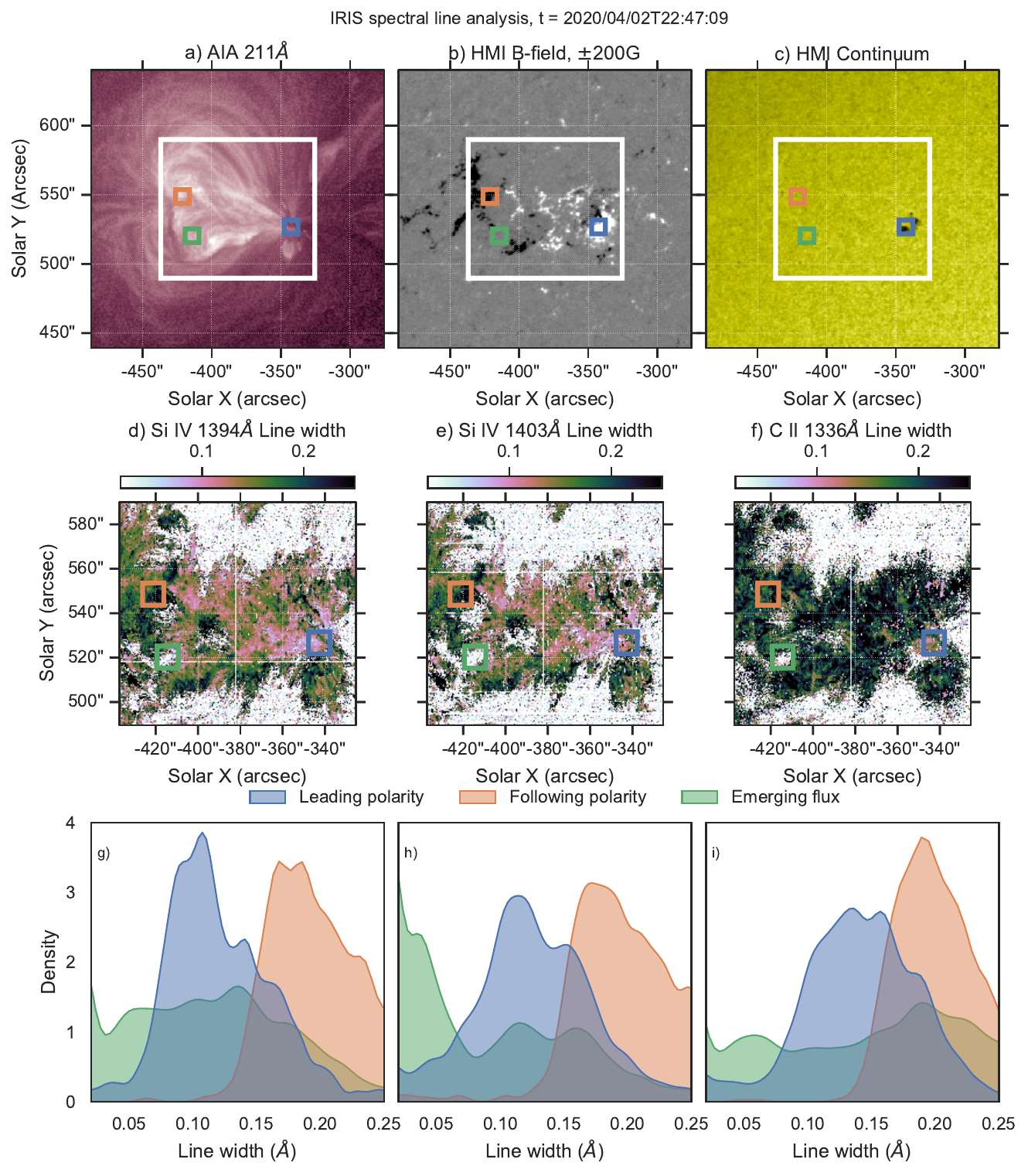}
    \caption{Plasma properties of AR~12759 at 22:47:09~UT on 2020-April-02. Panels a, b, \& c show the AIA 211~\AA, HMI line-of-sight magnetic field, and HMI continuum respectively. The white box shows the part of the IRIS raster field-of-view roughly corresponding to the core of the active region which was the primary focus of this work, with the coloured boxes showing the leading polarity (blue), following polarity (orange), and emerging flux (green) regions. Middle row shows the line width, and bottom row shows kernel density estimator (KDE) plots of the line width distributions in the three boxes for the \ion{Si}{4} 1394~\AA\ (left), \ion{Si}{4} 1403~\AA\ (middle), and \ion{C}{2} 1336~\AA\ (right) spectral lines. In each case, the images and plots have been limited to a range of 0.02-0.25~\AA. Note that flux emergence had not yet started in the ``emerging flux'' (green) region at the time of this figure.}
    \label{fig:IRIS_lines}
\end{figure*}

The IRIS observations of AR~12759 described here are outlined in Table~\ref{tbl:iris}. All of the rasters are very large, dense, 320-step rasters rebinned $2\times2$ with a 9.2~s \corr{exposure time} and a raster step of $0.35''$, giving a total field of view of $112''\times175''$ (shown by the blue dashed box in Figure~\ref{fig:context}) with a raster cadence of $\sim$49~minutes. IRIS calibrated level 2 data were used for this analysis, with the data already corrected for dark current, flat field, and geometrical distortion \citep[see, e.g.,][for more details]{depontieu:2014}. The IRIS data were then aligned with SDO/AIA observations using \corr{the SDO/AIA 304~\AA\ passband and the 2796~\AA\ passband from} the associated IRIS slitjaw images. Detailed IRIS analysis was performed on a subfield region-of interest corresponding to the white box shown in Figure~\ref{fig:IRIS_lines}, which roughly corresponds to the core of the observed active region. Note that the only differences between rasters with OBSID 3610108077 and 3620108077 is the use of lossless (3610108077) vs.\ default (3620108077) compression.

The 17 separate \emph{Hinode}/EIS observations of AR~12759 between 2 and 7 April 2020 described here were taken using the HPW021\_VEL\_260x512v2 \corr{study}. This \corr{study} contains the \ion{Si}{10} 258.375~\AA, \ion{S}{10} 264.233~\AA, \ion{Ca}{14} 193.874~\AA, and \ion{Ar}{14} 194.396~\AA\ lines as well as a series of Fe lines, and has previously been used to study active region FIP bias \citep[e.g.,][]{Testa:2023,To:2023}. The HPW021\_VEL\_260x512v2 \corr{study} scans a field-of-view of $260''\times512''$ with 87 raster positions of 40~s exposure time and uses the $2''$ slit width with a raster step of $3''$. The data were reduced using the SSWIDL eis\_prep.pro routine, which removes pixels affected by cosmic ray hits, dust, and electric charge, and corrects the data for instrumental effects including orbital spectrum drift and CCD spatial offset. The FIP bias here was estimated using two distinct line ratio diagnostics, namely the \sis\ and \caar\ ratios. The FIP bias derived from the \sis\ diagnostic was calculated using the technique developed by \citet{Brooks:2015} and subsequently used by \citet{Baker:2021,To:2021,To:2023}, in which spectral lines from consecutive ionisation stages of \ion{Fe}{8} - \ion{Fe}{17} were fit with single or multiple Gaussians as appropriate (typically single Gaussians unless the line is blended). These lines have a formation temperature of $\sim$0.5--5.5~MK, and the diagnostic can be used to estimate the FIP bias assuming a density estimated using the \ion{Fe}{13} 202.04~\AA/203.83~\AA\ line ratio. The \ion{Ca}{14} and \ion{Ar}{14} lines have a higher formation temperature of $\sim$3.5~MK, and the diagnostic was derived by taking the ratio of the two lines as \caar\ \citep[cf.][]{Baker:2019}. An example of the \corr{resulting composition} maps derived for both ratios is shown in Figure~\ref{fig:context}a,b, with both plots scaled using the same FIP bias range for consistency.

\section{Analysis techniques}
\label{s:analysis}

Given the unknown signatures of the plasma fractionation process in the solar chromosphere and transition region, a number of different analysis techniques were applied to each IRIS raster. The \ion{Si}{4} 1394~\AA, \ion{Si}{4} 1403~\AA, \ion{C}{2} 1336~\AA, and \ion{Mg}{2} k \& h spectral lines observed by the IRIS spectrograph are the main focus of this work as they provide an overview of plasma processes occurring through the chromosphere and transition region. Fortunately, all of the rasters used here (see Table~\ref{tbl:iris} for details) provided \corr{observations} of each of these spectral lines \corr{with high spectral resolution across a large field of view which encompassed a significant fraction of the EIS field-of-view with measureable FIP bias values}. 

It is clear from Table~\ref{tbl:iris} that the IRIS rasters were taken in a series of 5 distinct groups, with (in most cases) multiple raster scans per observation time. This allowed a detailed analysis of the active region within these particular observing windows. For brevity, the different analysis techniques are described in this section, with the evolution of different parameters within each group then discussed in Section~\ref{s:events}. 

\begin{deluxetable}{ccccc}
\tablecaption{Analysed IRIS observations of AR~12759.\label{tbl:iris}}
\tablehead{
\colhead{Start Time (UT)} & \colhead{OBSID} & \colhead{No.\ rasters} & \colhead{x, y} & \colhead{Group}
}
\startdata
2~Apr~22:47:09 & 3610108077 & 3 & $-386'', 534''$ & 1 \\
3~Apr~01:17:35 & 3610108077 & 3 & $-368'', 534''$ & 1 \\
3~Apr~04:48:09 & 3610108077 & 3 & $-356'', 535''$ & 1 \\
3~Apr~07:18:35 & 3610108077 & 3 & $-325'', 535''$ & 1 \\
3~Apr~09:49:21 & 3610108077 & 2 & $-309'', 537''$ & 1 \\
3~Apr~11:30:41 & 3610108077 & 1 & $-294'', 537''$ & 1 \\
3~Apr~20:16:44 & 3610108077 & 4 & $-233'', 542''$ & 2 \\
3~Apr~23:41:07 & 3610108077 & 3 & $-203'', 540''$ & 2 \\
4~Apr~02:11:53 & 3610108077 & 2 & $-183'', 541''$ & 2 \\
4~Apr~12:22:19 & 3620108077 & 4 & $-109'', 542''$ & 3 \\
4~Apr~15:46:16 & 3620108077 & 2 & $-79'', 543''$ & 3 \\
4~Apr~17:27:38 & 3620108077 & 4 & $-66'', 541''$ & 3 \\
4~Apr~20:46:55 & 3620108077 & 1 & $-40'', 544''$ & 3 \\
5~Apr~12:03:35 & 3620108077 & 3 & $76'', 542''$ & 4 \\
5~Apr~14:38:45 & 3620108077 & 5 & $98'', 543''$ & 4 \\
5~Apr~18:46:59 & 3620108077 & 4 & $130'', 542''$ & 4 \\
5~Apr~22:06:16 & 3620108077 & 3 & $158'', 541''$ & 4 \\
6~Apr~00:36:36 & 3620108077 & 4 & $177'', 542''$ & 4 \\
6~Apr~16:40:19 & 3620108077 & 5 & $297'', 542''$ & 5 \\
6~Apr~20:48:33 & 3620108077 & 3 & $328'', 539''$ & 5 \\
6~Apr~23:35:39 & 3620108077 & 4 & $344'', 537''$ & 5 \\
7~Apr~02:54:56 & 3620108077 & 1 & $369'', 534''$ & 5 \\
\enddata
\end{deluxetable}

Initial inspection of the magnetic evolution of the active region and the individual groups of IRIS rasters enabled identification of three regions-of-interest which are the focus of detailed analysis here. These regions correspond to the leading and following polarities of the active region, and a small emerging flux region which began emerging from $\sim$21:30~UT on 2020-April-03. Note that the regions-of-interest were defined separately for each group due to the small variability of position within each group, while also minimising variation in the location of these regions between the different groups. 

\subsection{Spectral line fitting}
\label{ss:spectra}

For each raster studied, the \ion{Si}{4} 1394~\AA, \ion{Si}{4} 1403~\AA, and \ion{C}{2} 1336~\AA\ lines were fitted using single Gaussians to derive the line intensity, width, and Doppler velocity. In each case, the lines were fitted using iris\_auto\_fit.pro from the SolarSoftWare \citep[SSW;][]{Freeland:1998} database. An initial inspection of the plasma parameters of each fitted line \corr{found that the intensity and Doppler velocity behaved as expected for an active region core, with no identifiable anomalous behaviour that could potentially be related to the measured EIS FIP bias. In contrast, the line width exhibited potentially interesting behaviour that required further inspection.} Figure~\ref{fig:IRIS_lines} shows the line width for the \ion{Si}{4} 1394~\AA, \ion{Si}{4} 1403~\AA, and \ion{C}{2} 1336~\AA\ lines at 22:47:09~UT on 2020-April-02 (panels~d-f), with the Kernel Density Estimation \citep[KDE; cf.][]{dejager:1986,Dacie:2016} plots in panels~g-i showing the distribution of pixel values within the boxes corresponding to the leading polarity (blue), following polarity (orange), and emerging flux (green) regions. Panels~a-c of the figure show the \corr{211}~\AA, HMI line-of-sight magnetic field, and HMI continuum for context, with the white box in each panel showing the field-of-view of the IRIS raster roughly corresponding to the core of the active region.

\begin{figure*}[!t]
    \includegraphics[width=0.99\textwidth]{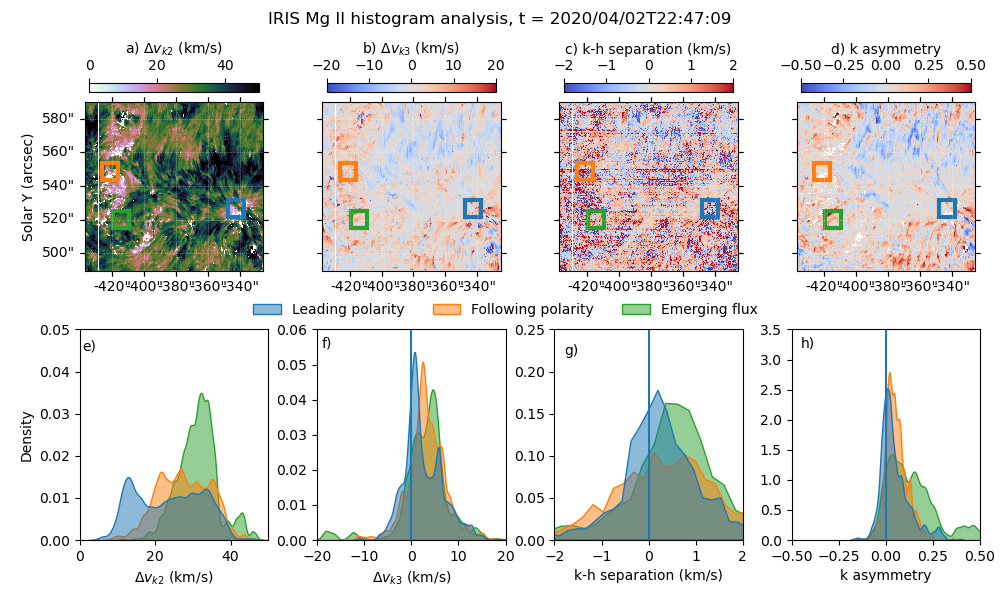}
    \caption{Fitted \ion{Mg}{2} properties of AR~12759 at 22:47:09~UT on 2020-April-02. Top row shows maps of (a) k2 separation, (b) k3 velocity, (c) k-h separation, and (d) k asymmetry, each calculated using iris\_get\_mg\_features\_lev2.pro. Bottom row shows the corresponding KDE plots of the pixel values within the regions-of-interest for each parameter.}
    \label{fig:IRIS_mgii}
\end{figure*}

It is clear \corr{from the KDE plots in panels g-i of Figure~\ref{fig:IRIS_lines}} that the \corr{line width values corresponding to the following polarity region are consistently higher than those corresponding to the leading polarity region, regardless of the line studied. In contrast}, the emerging flux region exhibits a much broader range of values \corr{typically peaking} at a lower value than observed for the following polarity region \corr{(with the exception here of the \ion{C}{2} line in panel~i)}. As noted by \citet{Martinez-Sykora:2023}, the \ion{Si}{4} line width provides an insight into unresolved velocity due to Alfv\'{e}n waves, so we chose the \ion{Si}{4} 1403~\AA\ line width for further analysis in 
Section~\ref{s:events}.

\subsection{Derivation of \ion{Mg}{2} properties}
\label{ss:mgii}

As discussed in detail by, e.g., \citet{Leenaarts:2013}, \citet{Pereira:2013} and \citet{Kerr:2015}, the \ion{Mg}{2} h \& k resonance lines are complex optically thick lines that form at multiple levels in the solar chromosphere. As a result, the lines are observed at different heights simultaneously, corresponding to the optical depth at a given frequency $\tau=1$, which allows the photons to escape. Both the h \& k lines have distinct shapes with features that can be used to probe different parts of the solar chromosphere \citep[e.g., Figure~1 of][]{Pereira:2013}. The k1 minima form near the temperature minimum, the k2 emission peaks form in the mid-chromosphere, while the k3 emission cores form in the upper chromosphere. The h \& k lines also form at slightly different heights in the solar chromosphere, with the k line forming a few tens of km higher due to its higher (by a factor of 2) opacity. The lines can also be used to probe temperature and velocity gradients between the formation heights of the k2r (red wing) and k2v (blue wing) components by measuring the asymmetry and separation of the emission peaks. To derive these properties, the \ion{Mg}{2} lines from each raster were analysed here using the SSWIDL iris\_get\_mg\_features\_lev2.pro routine \citep[cf. IRIS Technical Note 39 and][]{Pereira:2013}.

Figure~\ref{fig:IRIS_mgii} shows some of the fitted \ion{Mg}{2} line properties as derived using the iris\_get\_mg\_features\_lev2.pro routine for the raster beginning at 22:47:09~UT on 2020-April-02. The top row shows the corresponding images. Panel~a shows the k2 separation, which provide an estimation of the mid-chromospheric velocity gradient, panel~b shows the k3 velocity, providing an estimate of the upper chromosphere velocity, panel~c shows the k-h peak separation, which is sensitive to the upper chromospheric velocity gradient, and panel~d shows the k asymmetry, defined as,
\begin{equation}
    k_{\textrm{asym}} = \frac{I_{k2v} - I_{k2r}}{I_{k2v} + I_{k2r}}, 
\end{equation}
where $I_x$ is the intensity at the defined location $x$. This gives the sign of the velocity above the $\tau=1$ level. Note that the h line exhibits comparable behaviour to the k line for each parameter, and is therefore not shown here for brevity. The bottom row of Figure~\ref{fig:IRIS_mgii} shows the corresponding KDE plots giving the probability density of the values contained within the regions corresponding to the leading polarity, following polarity, and emerging flux. A reference line has also been added to the k3 velocity, k-h separation, and k asymmetry plots to indicate where the values equal 0.

For the raster shown in Figure~\ref{fig:IRIS_mgii}, the region where magnetic flux begins to emerge starting from $\sim$21:30~UT on 2020-April-03 has a higher k2 separation, indicating a higher mid-chromospheric velocity gradient than the following or leading polarities, both of which have very broad distributions. While all three regions have a comparable k3 (upper chromosphere) velocity and k-h separation (indicating comparable upper chromosphere velocity gradients), the emerging flux region does appear to have a slightly more positive upper chromosphere velocity gradient, although the actual flux emergence episode only starts about 23 hours later. Finally, all three regions have a comparable mainly positive k asymmetry, indicating mostly positive velocity above $\tau=1$.

\subsection{IRIS$^2$ inversion}
\label{ss:iris2}

The IRIS$^2$ inversion database of \citet{Sainzdalda:2022} was also used to gain an additional insight into the evolution of the chromosphere as AR~12759 transited the disc. The IRIS$^2$ inversions use a series of representative profiles, each with an associated Representative Model Atmosphere (RMA), where the RMA was derived using the STockholm inversion Code \citep[STiC;][]{delacruz:2019}. The spectral profile in each pixel of the individual rasters is compared to a lookup table of representative profiles, with the best fit returned. This provides an estimation of the turbulence velocity (v$_{turb}$), line-of-sight velocity (v$_{LOS}$), electron density (n$_{e}$), and temperature (T) with optical depth for each pixel in the rasters. As noted by \citet{Testa:2023}, the IRIS$^2$ inversions are optimised for optical depths in the range $-3.8 < \tau < -5$, so following their lead we use IRIS$^2$ images at $\tau = -4.2$ throughout this work.

\begin{figure*}[!t]
    \includegraphics[width=\textwidth]{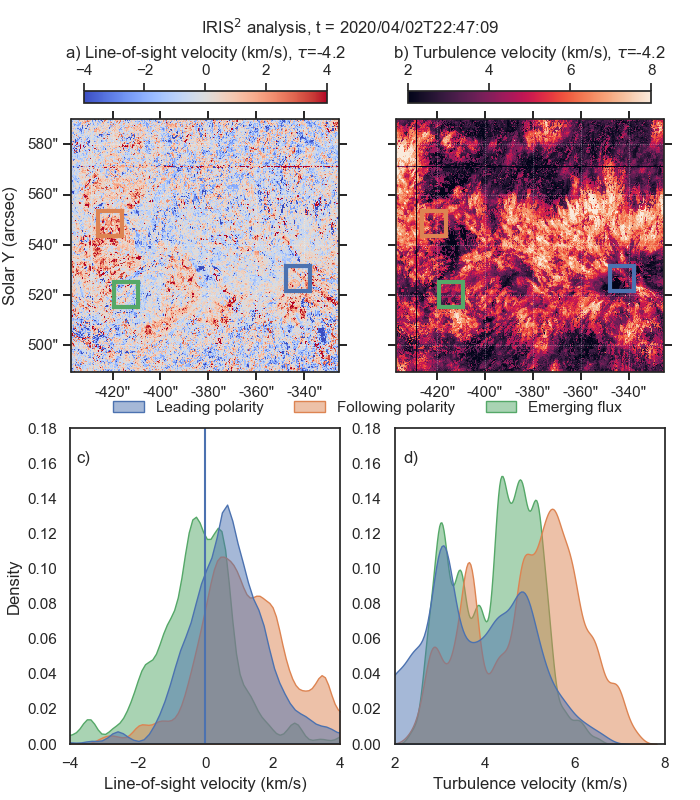}
    \caption{Derived IRIS$^2$ properties of AR~12759 at 22:47:09~UT on 2020-April-02. Top row shows maps of (a) line-of-sight velocity, and (b) turbulence velocity, both at an optical depth $\tau=-4.2$ \citep[cf.][]{Testa:2023}. Bottom row shows the corresponding KDE plots of the pixel values within the regions of interest for each parameter. Note that flux emergence had not yet started in the ``emerging flux'' (green) region at the time of this figure.}
    \label{fig:IRIS2}
\end{figure*}

The top row of Figure~\ref{fig:IRIS2} shows maps of the line-of-sight (a) and turbulence velocities (b) calculated using the IRIS$^{2}$ inversion for the raster starting at 22:47:09~UT on 2020-April-02. The bottom row shows the corresponding KDE plots of line-of-sight (c) and turbulence velocity (d) for the distribution of pixel values in the regions corresponding to the leading polarity, following polarity, and emerging flux. Again, a vertical reference line has been added showing the line-of-sight velocity equal to 0 to help guide the eye.

All three distributions of the line-of-sight velocity are quite broad, with the leading and following polarity regions predominantly positive, while the emerging flux region distribution is approximately symmetric about 0, with a slight bias towards the negative. All three regions have a spread in turbulence velocity from 2--7~km~s$^{-1}$, albeit each with two distinct peaks. 

\section{Analysis of individual groups}
\label{s:events}

\begin{figure*}[!t]
    \includegraphics[width=0.99\textwidth]{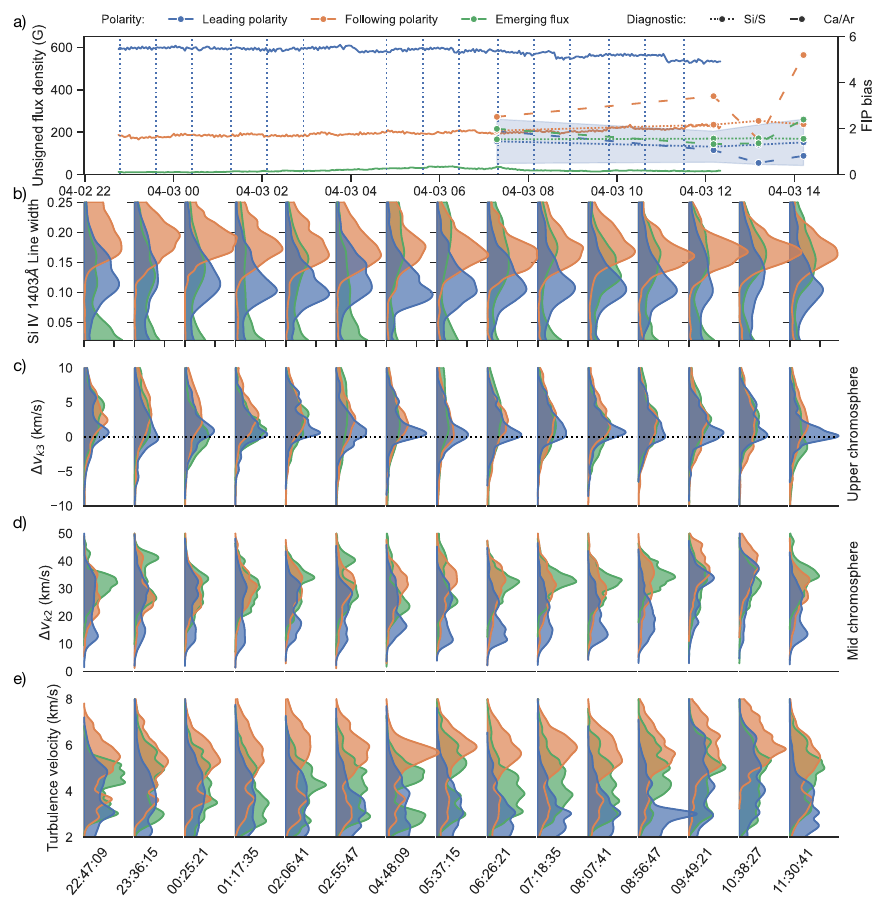}
    \caption{Panel (a) shows temporal evolution of SDO/HMI unsigned magnetic flux (solid lines) and mean Hinode/EIS FIP bias for the leading polarity (blue), following polarity (orange), and emerging flux (green) regions within the IRIS rasters identified as Group 1. FIP bias is calculated using the \ion{Si}{10}/\ion{S}{10} (dotted lines) and \ion{Ca}{14}/\ion{Ar}{14} (dashed lines) diagnostics. \corr{The blue shaded region shows the uncertainty associated with the \ion{Si}{10}/\ion{S}{10} ratio in the leading polarity region to provide a qualitative representation of the uncertainty associated with the FIP bias measurement.} Panels (b-e) show the temporal evolution of KDE plots of \ion{Si}{4}~1403~\AA\ line width (b), \corr{\ion{Mg}{2}} k3 velocity (c), \corr{\ion{Mg}{2}} k2 separation (d), and turbulence velocity calculated using the IRIS$^2$ inversions at $\tau=$-4.2 (e). In each case, colours show the different regions-of-interest corresponding to panel~a. Note that flux emergence had not yet started in the ``emerging flux'' (green) region at the time of this figure.}
    \label{fig:group1}
\end{figure*}

The different analysis techniques outlined in Section~\ref{s:analysis} were applied to each of the IRIS rasters across the entire observational period. Although not all of the different parameters exhibited any significant change or evolution with time, notable changes were identified for some of the parameters suggesting that further analysis was warranted. In particular, the line width for the \ion{Si}{4} 1403~\AA\ line was chosen for further analysis as there were significant differences in the observed behaviour between the different regions-of-interest. This parameter has also been used by \citet{Martinez-Sykora:2023} to infer the presence of unresolved Alfv\'{e}n waves. The turbulence velocity derived from the IRIS$^2$ inversions has previously been used by \citet{Testa:2023} to probe the fractionation process and relationship with FIP bias in outflow regions. \citet{Mihailescu:2023} have suggested that the ponderomotive force could be acting in the mid- or upper-chromosphere depending on the presence of resonant vs.\ non-resonant waves. This is consistent with the suggestion of \citet{Martinez-Sykora:2023} that in the normal collisional environment of the chromosphere, if the ponderomotive force is the dominant force in fractionation, waves should propagate from the chromosphere upward \citep[a suggestion supported by observations made by][]{Murabito:2023}. We also therefore analysed the k2 separation ($\Delta v_{k2}$) and k3 velocity ($\Delta v_{k3}$), which can be used to probe the velocity in the mid- and upper chromosphere respectively \citep{Leenaarts:2013,Pereira:2013}.

\subsection{Group 1}
\label{ss:grp1}

\begin{figure*}[!t]
    \includegraphics[width=0.99\textwidth]{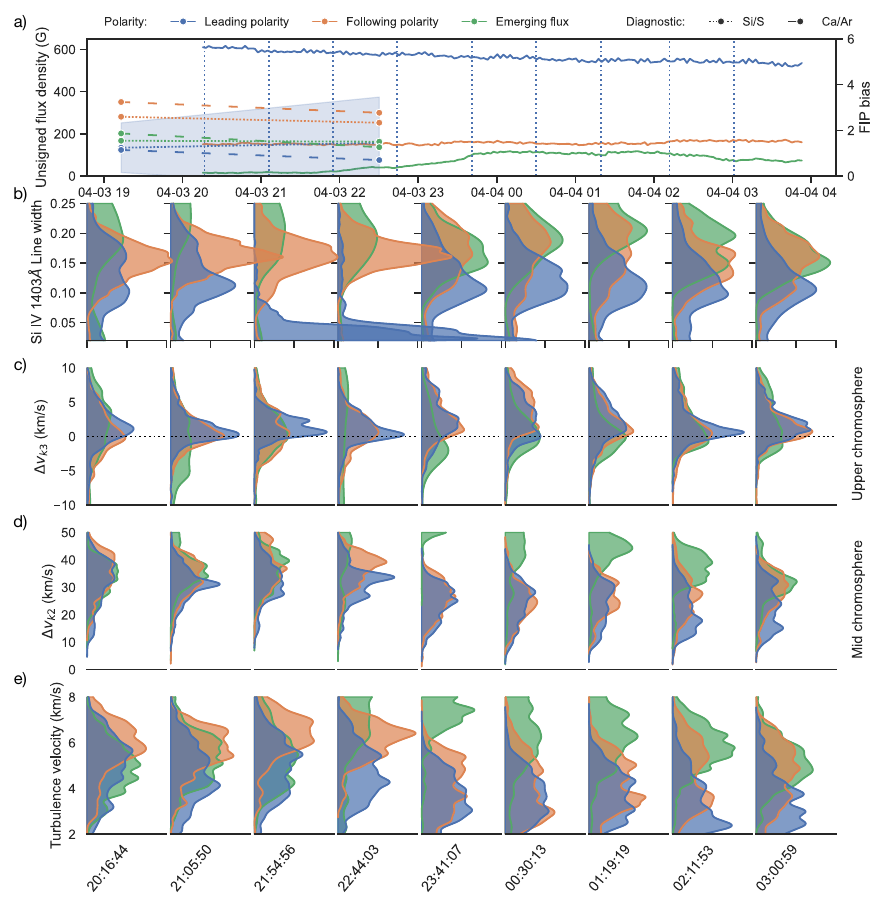}
    \caption{As Figure~\ref{fig:group1}, but for IRIS Group 2. Note that flux emergence begins in the ``emerging flux'' (green) region at approximately 21:30~UT during this sequence.}
    \label{fig:group2}
\end{figure*}

Group 1 covered a time period from 22:45~UT on 2020-April-02 until 12:20~UT on 2020-April-03, and included 15 IRIS rasters across 6 separate pointings. The active region was relatively quiet over this period, with no flares observed and little if any change in the different parameters shown in Figures~\ref{fig:IRIS_lines} - \ref{fig:IRIS2}. This can be seen in Figure~\ref{fig:group1}, which shows the evolution in HMI unsigned magnetic flux \corr{density (G)} and mean \emph{Hinode}/EIS FIP bias estimated using the \sis\ (dotted line) and \caar\ (dashed line) diagnostics (panel a) and the temporal evolution of the KDE plots for the \ion{Si}{4} 1403~\AA\ line width (panel~b), k2 separation (panel~c), k3 velocity (panel~d), and turbulence velocity (panel~e), in each case with the different colours corresponding to the three identified regions-of-interest. 

The unsigned magnetic flux \corr{density} exhibits no distinct changes in this time period, with the leading polarity region having the highest unsigned flux \corr{density}, followed by the following and emerging flux region. Note that the flux in this region had not yet begun to emerge by this time, so this is effectively a quiet Sun region. Although there are no \emph{Hinode}/EIS observations associated with this IRIS group prior to 07:00~UT on April 3, there is a distinct difference between the two FIP bias diagnostics estimated in the different regions-of-interest. There is a clear separation between the \sis\ and \caar\ estimates in the different regions-of-interest, with the following polarity region having the highest FIP bias value, followed by the emerging flux region, then the leading polarity region. It is also interesting that while the \sis\ diagnostic values are quite close for each of the three regions of interest (clustered about $\sim$2), the \caar\ diagnostic values are much more separated (clustered about $\sim$1 for the leading polarity and emerging flux regions and up to 
\corr{a maximum of 5} for the following polarity region). 

\begin{figure*}[!t]
    \includegraphics[width=0.99\textwidth]{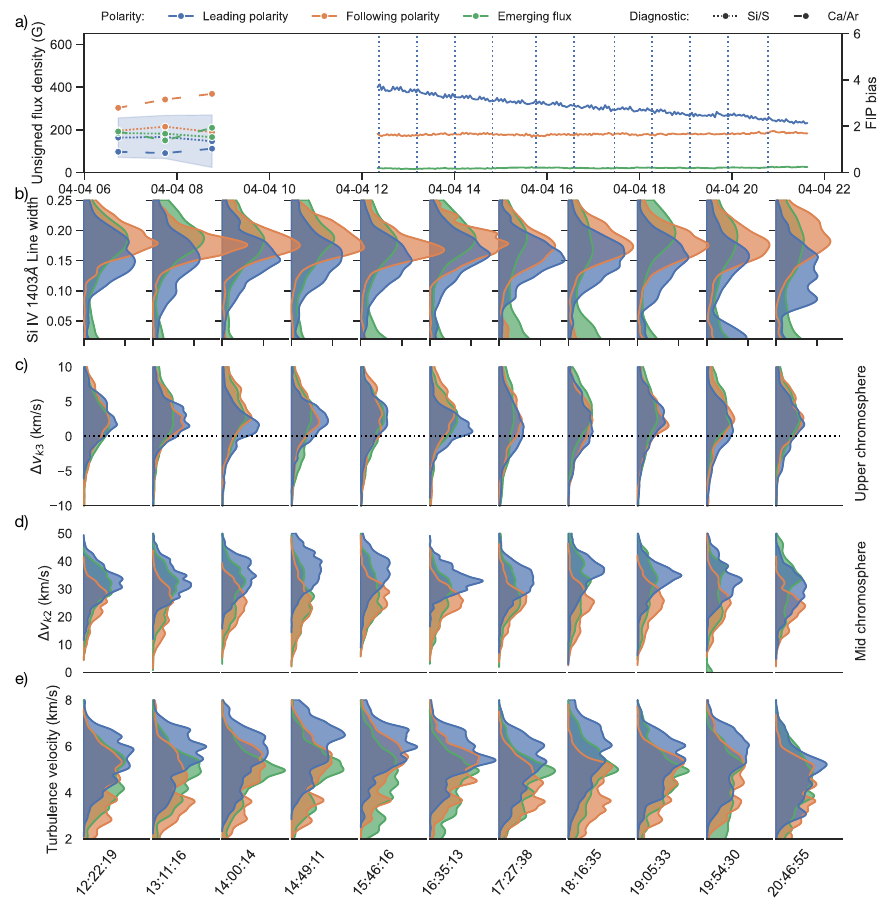}
    \caption{As Figure~\ref{fig:group1}, but for IRIS Group 3.}
    \label{fig:group3}
\end{figure*}

There is a distinct difference in \ion{Si}{4} line width between the different regions-of-interest, with the leading polarity region having a broad distribution peaking at lower values than the narrower distribution for the following polarity region. In contrast, the emerging flux region tends to have a broad distribution with two peaks, at both small and large values. Each of the three regions exhibit a broadly positive k3 velocity, indicating upward-directed velocity in the upper chromosphere \citep[a height of $\sim$2.5~Mm;][]{Pereira:2013}. However, the distribution associated with the leading polarity shows a strong peak at $\Delta v_{k3}=0$, with a long positive tail, while the following polarity and emerging flux regions both peak at positive velocities. For the k2 separation, corresponding to the mid-chromosphere velocity \citep[a height of $\sim$1.5~Mm;][]{Pereira:2013}, the velocity distributions in each case are quite broad. The emerging flux and following polarity regions tend to have distributions peaking at higher values than the leading polarity region, which tends to have quite a low peak value. It is notable that while the three regions exhibited relatively broad turbulence velocity distributions, the distributions tended to have higher values in the following polarity region than the leading polarity region, with the emerging flux region distribution typically falling in the middle. This broadly matches the behaviour of the FIP bias values.

\subsection{Group 2}
\label{ss:grp2}

Group 2 covered a time period from 20:15~UT on 2020-April-03 until 04:00~UT on 2020-April-04, including 9 IRIS rasters across 3 different pointings. This period was most notable for a small emerging flux region which emerged in the southeast of the active region, beginning at $\sim$22:30~UT (see top panel of Figure~\ref{fig:group2}). 

The emergence of magnetic flux was associated with an increase in the $\Delta v_{k2}$ (Figure~\ref{fig:group2}d) and turbulence velocity (Figure~\ref{fig:group2}e) measured in this region. However, although the $\Delta v_{k3}$ distribution measured in this region jumps to high positive values at $\sim$22:44~UT, it then becomes a skewed distribution peaking at 0 for subsequent rasters. This suggests that the effects of the flux emergence are primarily limited to the mid chromosphere and microturbulence measurements. While the leading and following polarity $\Delta v_{k2}$ and $\Delta v_{k3}$ distributions exhibit comparable behaviour, the distribution of turbulence velocity had typically higher values for the following polarity region than the leading polarity region. The \ion{Si}{4} line width distribution (Figure~\ref{fig:group2}b) in the emerging flux region does not exhibit any behaviour that can be directly related to the flux emergence, while the leading and following polarity distributions behave similarly to Group~1, with the distribution for the following polarity peaking at higher line width values than the distribution for the leading polarity, which peaks much closer to zero.

\begin{figure*}[!t]
    \includegraphics[width=0.99\textwidth]{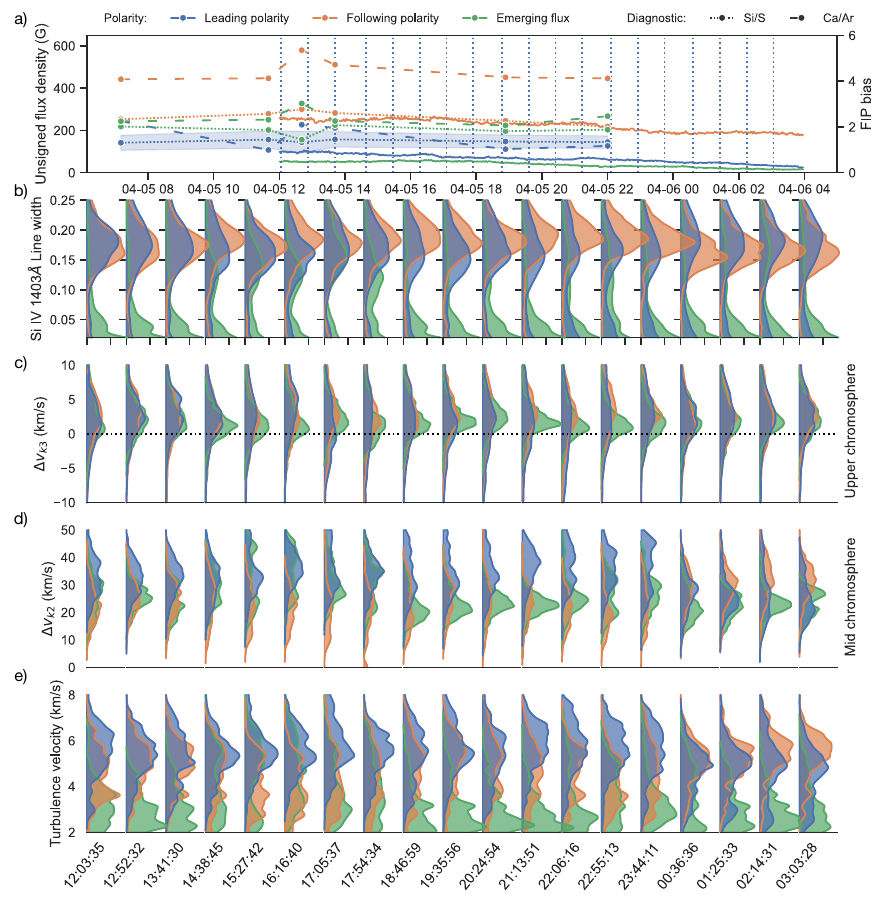}
    \caption{As Figure~\ref{fig:group1}, but for IRIS Group 4.}
    \label{fig:group4}
\end{figure*}

Unfortunately there were only two EIS rasters associated with this group, with both prior to the flux emergence, so no clear change in FIP bias value could be identified as a result of this flux emergence. However, there is a clear similarity in FIP bias measurements between the \sis\ and \caar\ diagnostics, with the values estimated in each region of interest comparable for each diagnostic.

\subsection{Group 3}
\label{ss:grp3}

Group 3 covered a time period from 12:20~UT until 21:45~UT on 2020-April-04, and included 11 IRIS rasters across 4 different pointings. This period was relatively quiet, with the unsigned flux \corr{density} associated with the leading polarity region exhibiting a clear decrease across the observing period (see Figure~\ref{fig:group3}a). The closest \emph{Hinode}/EIS observations of AR~12759 associated with this group occurred $\sim$4~hours prior to the first IRIS raster, and as with Groups 1 \& 2, the following polarity showed the highest FIP bias values, followed by the emerging flux region, and then the leading polarity region. As with Group~2, the leading polarity and emerging flux regions exhibited comparable FIP bias values for both the \sis\ and \caar\ diagnostics. However, the following polarity region has a separation between values for both diagnostics, similar to Group~1.

\begin{figure*}[!t]
    \includegraphics[width=0.99\textwidth]{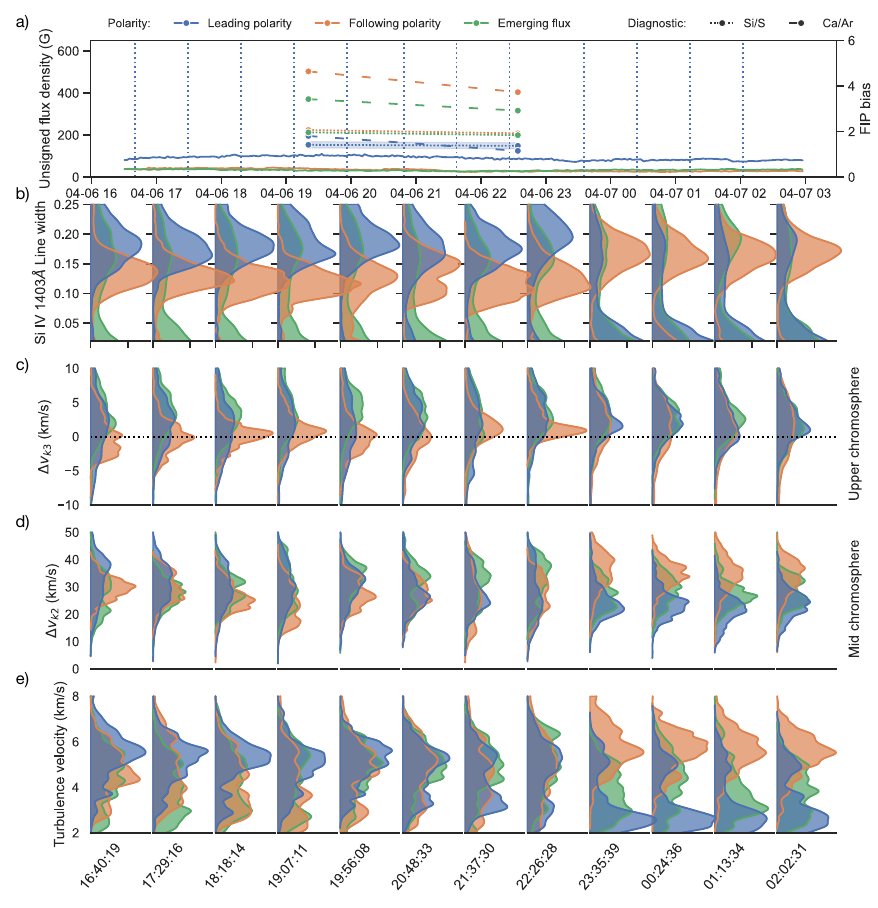}
    \caption{As Figure~\ref{fig:group1}, but for IRIS Group 5.}
    \label{fig:group5}
\end{figure*}

While the distributions of \ion{Si}{4} line width are comparable to the previous groups 1 \& 2 (Figure~\ref{fig:group3}b), the distribution for the leading polarity peaks at a higher value that is closer to that of the following polarity. The emerging flux region distribution is also comparable, although again does tend to have a second peak much closer to zero. The distributions of $\Delta v_{k3}$ (Figure~\ref{fig:group3}c) are broadly comparable between the three regions of interest, mainly positive with a peak close to 0. However, the leading polarity and emerging flux $\Delta v_{k2}$ distributions (Figure~\ref{fig:group3}d) peak at a higher positive value than the following polarity. The turbulence velocities (Figure~\ref{fig:group3}e) exhibit similar behaviour, with higher values observed in the leading polarity region. It is clear from the top panel of Figure~\ref{fig:group3} that the unsigned flux associated with the leading polarity is dropping significantly during this period. However, the magnetic flux which emerged as observed in Group~2 (Section~\ref{ss:grp2}), does merge with the following polarity region. This increased concentration of the magnetic field could explain the lower observed turbulence and $\Delta v_{k2}$ velocities in this region.

\subsection{Group 4}
\label{ss:grp4}

Group 4 covered a time period from 12:00~UT on 2020-April-05 until 04:00~UT on 2020-April-06, and included 19 IRIS rasters across 5 different pointings. By this stage, the sunspot in the leading polarity region had completely disappeared (at $\sim$01:00~UT on 2020-April-05). A filament had also begun to form along the inversion line of the bipole at $\sim$03:00~UT on 2020-April-05. By now, the unsigned flux \corr{density} of the leading polarity region is comparable to the quiet Sun (as shown by the ``emerging flux" polarity in Figure~\ref{fig:group4}a). 

The behaviour and evolution of each of the FIP bias diagnostics remains consistent in the three regions-of-interest throughout the time period of this grouping. Once again there is a clear separation between the FIP bias values for the \sis\ and \caar\ diagnostics in the following polarity region, with values of $\sim$4 ($\sim$2) in the \caar\ (\sis) diagnostic. The leading polarity and ``emerging flux'' regions exhibit broadly comparable values.

The distributions of \ion{Si}{4} line width in the leading and following polarity regions (Figure~\ref{fig:group4}b) have broadly comparable peaks, although the distribution is much narrower for the following polarity region than for the leading polarity region. The ``emerging flux'' region now peaks very close to zero, consistent with it now becoming part of the the background quiet Sun.

Similarly, the ``emerging flux'' $\Delta v_{k3}$ (Figure~\ref{fig:group4}c), $\Delta v_{k2}$ (Figure~\ref{fig:group4}d), and turbulence velocity (Figure~\ref{fig:group4}e) distributions all exhibit low velocity values, again consistent with this region having effectively become background quiet Sun by this stage of its evolution. Although the leading and following polarities have comparable $\Delta v_{k3}$ distributions, there are clear differences between the leading and following polarity distributions for the $\Delta v_{k2}$ and turbulence velocities, with the leading polarity exhibiting consistently higher values. This suggests that the magnetic field in the following polarity region is too weak to support observable activity in the upper chromosphere. 

\subsection{Group 5}
\label{ss:grp5}

Group 5 covered a time period from 16:40~UT on 2020-April-06 until 03:00~UT on 2020-April-07 and included 12 IRIS rasters across 3 different pointings. At this point, AR~12759 was approaching the West limb and had mostly dispersed magnetic field (as shown by the low unsigned magnetic flux \corr{density} in the top panel of Figure~\ref{fig:group5}) and a clear filament along its polarity inversion line. 

Only two \emph{Hinode}/EIS rasters were associated with this group of IRIS rasters, with the regions-of-interest exhibiting FIP bias values and behaviour comparable to that seen for Groups~1-4. As before, there is a separation between the \sis\ and \caar\ values, particularly in the following polarity region, but also now in the leading polarity region.

The \ion{Si}{4} line width distribution in the leading polarity region is now peaking at a higher value than the distribution in the following polarity region, although this behaviour then disappears from $\sim$23:35~UT with the leading polarity distribution peaking close to zero. The $\Delta v_{k3}$ and $\Delta v_{k2}$ distributions for the following polarity are centred around 0 and $\sim$30~km~s$^{-1}$ until $\sim$23:35:39~UT at which point they both broaden. The turbulence velocity also starts to increase at this point. This is most likely related to the onset of the filament eruption, observed at $\sim$01:30~UT. The leading polarity and emerging flux distributions are comparable in each case, with the leading polarity distribution values dropping towards the onset of the filament eruption. Note that at this point the decaying active region is also approaching the limb, which may affect the observed distributions.

\section{Discussion}
\label{s:disc}

The best current model to understand the FIP effect has been proposed by \citet{Laming:2004,Laming:2009,Laming:2015}, and uses the ponderomotive force to explain observations. In this model, Alfv\'{e}n waves originating in the corona induce a ponderomotive force when they refract and reflect at the high density gradient at the top of the chromosphere, which carries easily ionised low-FIP elements into the corona producing the observed FIP effect as they travel back and forth between the loop's adjacent footpoints. While this process should occur near the $\beta=1$ layer in the chromosphere/transition region, it has not yet been directly observed. Although the IRIS spacecraft regularly observes the chromosphere/transition region and could therefore provide the missing link in these studies, it does not observe spectral lines suitable for directly measuring the FIP effect \citep[cf.][]{depontieu:2021}.

Despite its inability to directly measure FIP bias, IRIS observations have previously been probed for signatures of elemental fractionation and the FIP effect. Previous work by \citet{Testa:2023} suggested a relationship between turbulence velocity estimated using the IRIS$^{2}$ inversions and FIP bias values in coronal outflow regions. However, they found no clear increase in turbulence in a high-FIP bias area close to a sunspot, and therefore they speculated that the difference might either suggest different properties in the underlying fractionaction mechanisms in different solar features, or could be possibly due more to the nature of the outflow region than being a direct signature of the fractionation process. Very little other work has been done on identifying signatures of fractionation in IRIS observations, with most work focussing on coronal \citep[e.g.,][]{Baker:2015,Brooks:2015} or ground-based chromospheric \citep[e.g.][]{Stangalini:2021,Murabito:2023} observations of FIP bias evolution or associated Alfv\'{e}n waves. In spite of this, there has been some recent work updating the ponderomotive force model to try and explain observations of differing FIP bias values in different loop populations within the same active region \citep{Mihailescu:2023}. In this case, the suggestion is that fractionation by the ponderomotive force is being driven at different heights in the chromosphere by resonant or non-resonant waves, with resonant waves acting near the top of the chromosphere, producing a milder fractionation signature, while non-resonant waves act lower in the chromosphere and produce a much stronger fractionation signature. This implies that it may be possible to use IRIS data to identify observable differences between different parts of the chromosphere consistent with fractionation processes and these resonant/non-resonant waves.

The three regions of interest presented in Sections~\ref{s:analysis} and \ref{s:events} were chosen as previous work has suggested identifiable differences in observed FIP bias between leading and following polarities within active regions, and emerging flux regions \citep[e.g.,][]{Baker:2018,Mihailescu:2022,To:2023}. The evolution of the magnetic flux of AR~12759 during the time period studied here clearly shows that the leading polarity region decays and disperses with time, while the following polarity region (although initially more dispersed than the leading polarity regions) ultimately becomes more compact as it incorporates the emerged flux from the emerging flux region (although a sunspot is never formed). In general, the turbulence velocity distributions reflect the dispersive nature of the magnetic flux within the different regions. The more dispersed the magnetic flux within a region, the broader and more positive the values within the distribution of turbulence velocity values. Similarly, more compact regions tend to have narrower distributions of turbulence velocity values that peak much closer to zero. This makes intuitive sense when considering the plasma-$\beta$; a more compact magnetic flux region inhibits plasma motion resulting in a lower turbulence velocity, whereas there is more opportunity for plasma motion in a more dispersed magnetic flux region, leading to a higher turbulence velocity \citep[cf.][]{To:2023}.

How does this then relate to plasma fractionation and the FIP bias? \citet{Mihailescu:2022} noted that FIP bias increases with magnetic flux density in the region $\leq$200~G, with that trend stopping for regions $\geq$200~G. Here, the leading polarity region initially corresponds to a sunspot, with an unsigned flux \corr{density} $>$200~G until the end of the third group of IRIS rasters (approximately 22:00~UT on April~4; see Figure~\ref{fig:group3}). Throughout this time period, the leading polarity region has a consistently lower associated FIP bias than the following polarity region where the unsigned flux \corr{density} is consistently below 200~G. The emerging flux region also follows this hypothesis, with a FIP bias value between that of the leading and following polarity regions and a consistent unsigned flux \corr{density} value of $\sim$50--100~G. These observations are also comparable with the work of \citet{To:2023} who found a similar relationship between coronal abundance and magnetic flux density using observations of this active region from 3 and 7 April 2020. It is interesting that this is also consistent with the work of \citet{Martinez-Sykora:2023}, who suggested that sunspots with strong flux densities approach a ``collisionless'' case, where waves from the chromosphere could generate an IFIP effect, thus lowering the FIP bias associated with the region (as noted here) or even creating an observable IFIP bias.

In addition to the turbulence velocity, the distributions of $\Delta v_{k3}$ and $\Delta v_{k2}$ velocities and the line width of the \ion{Si}{4}~1403~\AA\ emission line within these regions-of-interest were tracked with time to identify any differences. The $\Delta v_{k3}$ and $\Delta v_{k2}$ velocities derived from the \ion{Mg}{2} lines can be used to probe the velocity in the upper- and mid-chromosphere respectively \citep[cf.][]{Pereira:2013,Leenaarts:2013}, the regions where resonant (upper chromosphere) and non-resonant (mid-chromosphere) waves should be acting. Similarly, the line width, and by extension the non-thermal velocity, provides an insight into unresolvable Alfv\'{e}n waves that could be the predicted resonant or non-resonant waves.

In Group~1 (Section~\ref{ss:grp1}), the leading polarity region exhibits a low mid-chromosphere velocity ($\Delta v_{k2}$), and a distribution in upper chromosphere velocity ($\Delta v_{k3}$) that peaks strongly at 0. It also has a broadly constant \sis\ and \caar\ FIP bias of $\sim$1, suggesting little-to-no fractionation, which is a characteristic value in sunspot umbrae \citep{Baker:2021}. In contrast, the following polarity and emerging flux regions have a higher mid- and upper-chromosphere velocity, with the mid-chromosphere velocity for the following polarity region tending to increase slightly with time. The following polarity region exhibits different \sis\ and \caar\ FIP bias values, with values of $\sim$2 (\sis) and $\sim$3-5 (\caar). Both the \sis\ and \caar\ FIP bias values in the emerging flux region also remain roughly constant, with a comparably constant mid-chromosphere velocity. The \ion{Si}{4} line width distribution \corr{consistently peaks} at a much higher value in the following polarity region than in the leading polarity region, \corr{indicating a higher non-thermal velocity and hence increased microturbulence, consistent with} increased wave activity in \corr{the following polarity region compared to the leading polarity} region.

The mid-chromosphere velocity then shows a significant jump during the period of flux emergence (as shown by the green distributions in Figure~\ref{fig:group2}), but the upper-chromosphere velocity is unaffected. Unfortunately there is no comparable FIP bias measurement at this time, although the two measurements taken prior to this flux emergence show that the two diagnostics are broadly comparable in the different regions-of-interest. This combination of chromospheric velocities suggests that the small-scale flux emergence, when it has just started, affected the lower solar atmosphere, but did not strongly affect the upper chromosphere, while no clear signature of the flux emergence is seen in the \ion{Si}{4} line width in the emerging flux region. The magnetic flux density in the leading polarity region then starts to decrease with time, (e.g., Figure~\ref{fig:group3}, although there are unfortunately no corresponding FIP bias measurements). As its previously compact magnetic flux disperses, the leading polarity region has increased mid- and upper-chromosphere velocity, and a \ion{Si}{4} line width distribution peaking at higher values, but no corresponding increase in either observable FIP bias diagnostic. In contrast, the following polarity region exhibits lower mid-chromosphere velocity, higher upper-chromosphere velocity, consistently high \ion{Si}{4} line width values, and a clear separation between the \sis\ and \caar\ diagnostics, with the \caar\ diagnostic consistently higher in the following polarity region.

As noted by \citet{Mihailescu:2023}, resonant waves fractionating plasma in the upper chromosphere should produce comparable FIP bias enhancements using both the \sis\ and \caar\ diagnostics, whereas non-resonant waves fractionating plasma in the lower chromosphere should produce significantly higher \caar\ values compared to \sis. The FIP bias diagnostic values presented here are consistent with non-resonant waves fractionating plasma in the following polarity region and resonant waves fractionating the plasma in the leading polarity and emerging flux regions. As a relatively simple bipolar active region with the leading and following polarities mostly connected to each other, this interpretation would appear to be rather anomalous. However, connectivity in a decaying active region is complex, so it is possible that magnetic field in the selected regions-of-interest were not connected to each other (in particular, the AIA \corr{211}~\AA\ image in Figure~\ref{fig:IRIS_lines}a shows that the following polarity is connected to a closer-by positive polarity than the ``leading polarity box''). As a result, while a broad connectivity between the two polarities of the active region is expected, it is indeed possible that the identified regions-of-interest here were dominated by either non-resonant or resonant waves. There has also been some work by \citet{Giannattasio:2013} noting an imbalance of velocity oscillations between the leading and following polarities of bipolar active regions. This could be a contributing factor to the observations described here, and requires further investigation. The associated IRIS observations are similarly complex, and require further analysis to fully interpret their relationship to the wave types predicted by the fractionation measurements. However, it is notable that the following polarity region exhibits a consistently higher \ion{Si}{4} line width\corr{, indicating a higher nonthermal width, consistent with} increased unresolved wave activity in this location compared to the leading polarity  and ``emerging flux'' regions.

\section{Conclusions}
\label{s:conc}

Fractionation of plasma in the solar atmosphere is a long observed and poorly understood process that has been the focus of significant research particularly since the launch of the \emph{Hinode} spacecraft, which has enabled spatially resolved observations of FIP bias. The leading model to explain this phenomenon, the ponderomotive force model, suggests that fractionation is driven by the ponderomotive force resulting from Alfv\'{e}n waves propagating into the chromosphere. This implies that it should be possible to observe some spectral evidence of this fractionation process in the chromosphere. Here we use five days of observations of a decaying active region from the IRIS spacecraft to try and identify any signatures of plasma fractionation in the solar chromosphere and/or transition region. 

A comparison of the FIP bias values estimated using the \emph{Hinode}/EIS \sis\ and \caar\ diagnostics in the three regions-of-interest found distinct differences between the two diagnostics in the following polarity region, but no clear differences in the leading polarity or ``emerging flux'' regions. These observations suggest weak (if any) fractionation in the leading polarity and emerging flux regions, and enhanced fractionation in the following polarity region. 

The clear differences between the FIP bias values estimated using both diagnostics in the leading and following polarity regions can be understood by examining their magnetic environment. The leading polarity region is home to a sunspot and has a high unsigned magnetic flux which drops with time as the sunspot decays. In contrast, the following polarity region has much more dispersed magnetic flux, and absorbs some of the emerging flux following its emergence and dispersal. This is consistent with the suggestion of \citet{To:2023} and \citet{Mihailescu:2022} of a connection between the magnetic flux and the degree of fractionation. 

The fractionation process should be occurring in the chromosphere, with \citet{Mihailescu:2023} suggesting that the ponderomotive force proposed by \citet{Laming:2004,Laming:2009,Laming:2015} as the driver of plasma fractionation should be induced in the 
upper or lower chromosphere if driven by resonant or non-resonant waves respectively. Despite a thorough analysis of IRIS observations of this region, no clear signature of this process could be identified here. However, a comparison of the \ion{Si}{4} line width distributions in the different regions-of-interest \corr{reveal clear disparities in the non-thermal broadening of the \ion{Si}{4} line in these regions, indicating variations in the turbulence velocity consistent with} increased unresolved wave activity in the following polarity region compared to the leading polarity region. The chromospheric velocities derived from the \ion{Mg}{2} lines also reveal some unusual behaviour, although a full interpretation requires predictions provided by modelling of the ponderomotive force model. Some of the observed and measured behaviour of the IRIS parameters do therefore warrant further investigation and analysis using a combination of observations and modelling, and we intend to follow this line of investigation in future work. 

\section*{Acknowledgements}
\corr{The authors wish to thank the anonymous referee whose suggestions helped to improve the paper.}
DML is grateful to the Science Technology and Facilities Council for the award of an Ernest Rutherford Fellowship (ST/R003246/1). 
DB is funded under Solar Orbiter EUI Operations grant number ST/X002012/1 and Hinode Ops Continuation 2022-25 grant number ST/X002063/1. 
LvDG acknowledges the Hungarian National Research, Development and Innovation Office grant OTKA K-131508. 
The work of DHB was performed under contract to the Naval Research Laboratory and was funded by the NASA Hinode program.
AWJ acknowledges funding from the STFC Consolidated Grant ST/W001004/1. 
PT was funded for this work by contracts 8100002705 (IRIS),
and NASA contract NNM07AB07C (Hinode/XRT) to the Smithsonian Astrophysical Observatory, and by the NASA Heliophysics grants 80NSSC21K0737, 80NSSC21K1684, and 80NSSC20K1272. 
MM has been supported by the ASI-INAF agreement n.~2022-14-HH.0 and by the Italian agreement ASI-INAF 2021-12-HH.0 “Missione Solar-C EUVST—Supporto scientifico di Fase B/C/D". 
IRIS is a NASA small explorer mission developed and operated by LMSAL with mission operations executed at NASA Ames Research Center and major contributions to downlink communications funded by ESA and the Norwegian Space Centre.
Hinode is a Japanese mission developed and launched by ISAS/JAXA, collaborating with NAOJ as a domestic partner, and NASA and STFC (UK) as international partners. Scientific operation of Hinode is performed by the Hinode science team organized at ISAS/JAXA. Support for the post-launch operation is provided by JAXA and NAOJ (Japan), STFC (UK), NASA, ESA, and NSC (Norway). AIA data courtesy of NASA/SDO and the AIA, EVE, and HMI science teams.

\facilities{SDO, IRIS, \emph{Hinode}/EIS}

\software{SolarSoftWare \citep{Freeland:1998}, Numpy \citep{Harris:2020}, Sunpy \citep{Sunpy:2020}, aiapy \citep{Barnes:2020}, EISPAC \citep{Weberg:2023}, Seaborn \citep{Waskom:2021}, Matplotlib \citep{Hunter:2007}, pandas \citep{McKinney:2010}}

\section*{Data Availability}
No new data were generated as part of this study. The IRIS data can be downloaded from the LMSAL IRIS website (\href{https://iris.lmsal.com/data.html}{https://iris.lmsal.com/data.html}). The Hinode EIS data can be downloaded from the Hinode Science Data Centre (\href{http://sdc.uio.no/sdc/}{http://sdc.uio.no/sdc/}).

\bibliography{bibliography}{}
\bibliographystyle{aasjournal}

\end{document}